\documentclass[twocolumn,showpacs,preprintnumbers,amsmath,amssymb,letterpaper]{revtex4}
\usepackage{graphicx}

\begin{document}
\sloppy

\title{Fidelity of holonomic quantum computations in the case of random errors in the values of control parameters}
\author{P. V. Buividovich}
\email{buividovich@tut.by}
\affiliation{Belarusian State University, 220080 Belarus, Minsk, F. Skoriny av. 4}
\author{V. I. Kuvshinov}%
\email{v.kuvshinov@sosny.bas-net.by}
\affiliation{JIPNR, National Academy of Science, 220109 Belarus, Minsk, Acad. Krasin str. 99}
\date{22 December 2005}
\begin{abstract}
We investigate the influence of random errors in external control parameters on the stability of holonomic quantum computation in the case of arbitrary loops and adiabatic connections. A simple expression is obtained for the case of small random uncorrelated errors. Due to universality  of mathematical description our results are valid for any physical system which can be described in terms of holonomies. Theoretical results are confirmed by numerical simulations.
\end{abstract}
\maketitle

\section{\label{sec:Introduction} Introduction} 

The idea of holonomic quantum computations (HQC) was mainly developed in the work \cite{Zanardi:99}. In holonomic quantum computer quantum gates are implemented as holonomies (parallel transporters along loops) in the space of external control parameters $\lambda^{\mu}$ \cite{Zanardi:99, Pachos:02, Pachos:05}. Practical implementations of such computational scheme \cite{Pachos:00, Pachos:02} are based on the systems described by some degenerate hamiltonian which depends on the set of external control parameters $\lambda^{\mu}$ \cite{Zanardi:99, Pachos:02, Pachos:05}:
\begin{eqnarray}
\label{HQCHamiltonian}
\hat{H} (\lambda) = \hat{U}(\lambda) \hat{H}_{0} \hat{U}^{\dag}(\lambda)
\end{eqnarray}
where $\hat{U}(\lambda)$ are some unitary operators. The holonomies are obtained in this case by adiabatically changing the values of the control parameters in such a way that the initial and the final set of parameters coincide (thus one passes along the loop in parametric space). The states of the quantum register are supposed to be the eigenvectors of the hamiltonian corresponding to a single degenerate energy level \cite{Zanardi:99, Pachos:00, Kuzmin:03}.

However practical implementations of HQC encounter many difficulties, which are caused either by decoherence processes or imperfect external control. While decoherence processes in quantum information processing is a well-studied subject, the influence of errors in the values of external control parameters on the stability of HQC is specific to this particular implementation of quantum computations and attracted less attention. Improved insensitivity of HQC to external perturbations is believed to be one of its main advantages, therefore it is important to study the stability of HQC in as general case as possible. Some works on the topic are \cite{Carollo:03, DeChiara:03, Solinas:05, Kuzmin:05, Kuzmin:03}. Geometric phase in open systems was studied, for instance, in \cite{Carollo:03}, where the geometric phase was shown to be insensitive to dephasing. The effect of fluctuations in the classical control parameters on the Berry phase of a spin 1/2 interacting with  magnetic field was analyzed in \cite{DeChiara:03}. Numerical study of the stability of a particular implementation of HQC with respect to parametric noise was recently presented in the paper \cite{Solinas:05}. Analytic expressions for the fidelity of HQC implementation of Hadamard gate were obtained in \cite{Kuzmin:05}, revealing improved robustness against perturbations up to the fourth order in the magnitude of errors. Non-abelian Stokes theorem was applied to study the stability of HQC in the case of arbitrary loops and adiabatic connections in the work \cite{Kuzmin:03}, but the expressions obtained in this work do not take into account randomness of errors.

In this work we will investigate the stability of HQC with respect to random errors in the values of external control parameters in the case of arbitrary loops  and arbitrary adiabatic connections. The paper is organized as follows: in this section we introduce some notations which will be convenient for further calculations, in the section \ref{sec:Fidelity} we follow the work \cite{Kuzmin:03} and use the non-Abelian Stokes theorem to obtain the expression for the fidelity of HQC. The section \ref{sec:RandomErrors} contains the main result of this paper, namely the expression for the fidelity in the case of small random errors. Finally, in the section \ref{sec:Simulations} theoretical results are compared with results of numerical simulations.

We will assume that the Hilbert space of the states of the quantum register used in HQC is $N$-dimensional. In order to describe the evolution of the state vector of the quantum register one needs to introduce the adiabatic connection $\hat{A}_{\mu}$ which has the following matrix structure \cite{Wilczek:84, Zanardi:99}:
\begin{eqnarray}
\label{AdiabaticConnectionDefintion}
    \left(\hat{A}_{\mu}\right)_{mn} (\lambda) = \langle \phi_{m} (\lambda) |
 - i \frac{\partial}{\partial \lambda^{\mu}} | \phi_{n} (\lambda) \rangle  
\end{eqnarray}
where $| \phi_{m} (\lambda) \rangle$, $m = 1 \ldots N$ are linearly independent state vectors of the quantum register.

An arbitrary quantum gate is represented as the holonomy of the adiabatic connection which corresponds to some loop $\gamma$ in the space of external control parameters \cite{Wilczek:84, Zanardi:99}:
\begin{eqnarray}
\label{HQCGATE} 
\hat{\Gamma}_{\gamma} = \mathcal{P} \exp \left( i \int \limits_{\gamma} d\lambda^{\mu} \hat{A}_{\mu} \left(\lambda
\right)
 \right)
\end{eqnarray}
where $\mathcal{P}$ is the path-ordering operator. By choosing an appropriate loop $\gamma$ one implements different unitary operators (quantum gates) acting on the state vector of the quantum register: $| \phi \rangle = \hat{\Gamma}_{\gamma} | \phi_{in} \rangle$, where $| \phi_{in} \rangle$ is the initial state of the quantum register. It should be noted that due to the mathematical universality of holonomy all our results can be actually applied to any physical system which can be described in terms of holonomies (the most interesting example is probably Yang-Mills field).

One can also state that evolution of the state vector of the quantum register is described by the parallel transport equation, which is equivalent to (\ref{HQCGATE}):
\begin{eqnarray}
\label{StateParallelTransport}
\frac{d}{d s} \: | \phi \rangle = i \tau^{\mu} \hat{A}_{\mu} | \phi \rangle
\end{eqnarray}
where $s$ is some parametrization of the path $\gamma$ and $\tau^{\mu} = \frac{d \lambda^{\mu}}{d s} $ is the tangent vector.

Let us introduce the density matrix $\hat{\rho} (\lambda) = | \phi (\lambda) \rangle \langle \phi (\lambda) | $ of the quantum register for further convenience. Parallel transport of the density matrix is then described by the following equation:
\begin{eqnarray}
\label{DMatrixParallelTransport}
\frac{d}{d s} \: \hat{\rho} = i \tau^{\mu} \left[ \hat{A}_{\mu}, \hat{\rho} \right]
\end{eqnarray}

It is convenient now to use the generators $\hat{T}_{a}$, $a = 1 \ldots N^{2} - 1$ of $SU(N)$ group, which are normalized as ${\rm Tr} \: \left(\hat{T}_{a} \hat{T}_{b} \right) = \delta_{a b}$. We use latin indices to denote the elements of $su(N)$ Lie algebra and greek indices for tensors on the manifold of external control parameters. The generators  $\hat{T}_{a}$ in fundamental representation build a basis in the space of traceless hermitan operators acting on $N$-dimensional Hilbert space, therefore we can decompose the adiabatic connection $\hat{A}_{\mu}$ and the density matrix $\hat{\rho}$ as:
\begin{eqnarray}
\label{LieAlgebraDecompositions}
    \hat{\rho}    =  N^{-1} \hat{I} +  \hat{T}_{a} \rho^{a}, \: 
    \hat{A}_{\mu} =  A^{0}_{\mu} \hat{I} +  \hat{T}_{a} A^{a}_{\mu}, \nonumber \\
    \rho^{a} =  {\rm Tr} \left( \hat{\rho} \hat{T}_{a} \right), \:
    A^{a}_{\mu} =  {\rm Tr} \left( \hat{A}_{\mu} \hat{T}_{a} \right), \:
    A^{0}_{\mu} = N^{-1} {\rm Tr} \: \hat{A}_{\mu}
\end{eqnarray} 
where we have taken into account that ${\rm Tr} \: \hat{\rho} = 1$ and assumed summation over double indices.
Equation (\ref{DMatrixParallelTransport}) can be rewritten in terms of $\rho^{a}$ and $A^{a}_{\mu}$ as:
\begin{eqnarray}
\label{DMatrixParallelTransportAdjoint}
\frac{d}{d s} \: \rho^{a} = \tau^{\mu} C^{a}_{\: bc} A^{b}_{\mu} \rho^{c}
\end{eqnarray}
where $C^{a}_{\: bc}$ are the structure constants of $su(N)$ Lie algebra: $\left[ \hat{T}_{b}, \hat{T}_{c} \right] = - i C^{a}_{\: bc} \hat{T}_{a}$. After omitting the indices of $\rho^{a}$ and introducing the adiabatic connection in the adjoint representation of $SU(N)$ group $\left(\hat{A}^{adj}_{\mu} \right){}^{a}_{c} = C^{a}_{\: bc} A^{b}_{\mu}$, $\hat{A}^{adj \: \dag}_{\mu} = - \hat{A}^{adj}_{\mu}$ we can also rewrite the equation (\ref{DMatrixParallelTransportAdjoint}) as:
\begin{eqnarray}
\label{DMatrixParallelTransportAdjointCompact}
\frac{d}{d s} \: \rho = \tau^{\mu} \hat{A}^{adj}_{\mu} \rho
\end{eqnarray}
We will also use the notation $\rho(\gamma)$, or $\hat{\rho}(\gamma)$ to denote the decomposition coefficients (\ref{LieAlgebraDecompositions}) or the density matrix obtained as a result of parallel transport along the loop $\gamma$. According to the equation (\ref{DMatrixParallelTransportAdjointCompact}) $\rho(\gamma)$ can be represented as:
\begin{eqnarray}
\label{DMatrixAdjointSolution}
\rho (\gamma) = \mathcal{P} \exp \left( \int \limits_{\gamma} d\lambda^{\mu} \hat{A}^{adj}_{\mu} \right) \rho_{in}
\end{eqnarray}
where $\rho_{in}$ is the initial density matrix of the quantum register: $\hat{\rho}_{in} = | \phi_{in} \rangle \langle \phi_{in} | = N^{-1} \hat{I} +  \hat{T}_{a} \rho^{a}_{in}$, $\rho^{a}_{in} = \langle \phi_{in} | \hat{T}_{a} | \phi_{in} \rangle$.

\section{\label{sec:Fidelity}Fidelity of holonomic quantum computer}

The values of external control parameters can always contain unpredictable errors, therefore it is necessary to know how these errors can change the results of quantum computations. The influence of errors in the values of the external control parameters on the stability of quantum computations was studied, for instance, in the works \cite{Kuzmin:03, Solinas:05}. In order to measure deviations of the results of quantum computations from the desired result we will use the quantity named the fidelity, which is conventionally used to characterize the stability of quantum systems with respect to external perturbations \cite{Peres:84, Kuzmin:03, Prozen:01, Solinas:05}. Exponential decay of the fidelity means instability of quantum system \cite{Peres:84}. Following the works \cite{Peres:84, Prozen:01} we define the fidelity as:
\begin{eqnarray}
\label{FidelityDefinition}
f = {\rm Tr} \: \left( \hat{\rho} (\gamma') \hat{\rho} (\gamma) \right)
\end{eqnarray} 
where $\hat{\rho}(\gamma')$ is the density matrix of the quantum register obtained as a result of parallel transport along the loop with errors $\gamma'$ and $\hat{\rho} (\gamma)$ is the density matrix which corresponds to the desired operation. The physical meaning of the fidelity is the probability of obtaining the correct result of quantum computations. We will assume here that the initial points of the loops $\gamma'$ and $\gamma$ coincide, as it is usually true in practical implementations \cite{Pachos:00, Pachos:02}. It is also convenient to express the fidelity in terms of the coefficients $\rho^{a}$ in the decomposition (\ref{LieAlgebraDecompositions}): 
\begin{eqnarray}
\label{FidelityAdjoint}
    f = N^{-1} + \rho^{a}(\gamma') \rho^{a}(\gamma) 
\end{eqnarray}

Using the expressions (\ref{DMatrixAdjointSolution}) and (\ref{FidelityAdjoint}) and taking into account the properties of path-ordered exponents we obtain for the fidelity:
\begin{eqnarray}
\label{FidelityAsLoopIntegral}
    f = N^{-1} + \rho^{b}_{in} 
    \mathcal{P} \exp \left( \int \limits_{\gamma'} dx^{\mu} \hat{A}^{adj}_{\mu} \right) {}^{c}_{b} \:  \nonumber \\ \times
    \mathcal{P} \exp \left( \int \limits_{\gamma} dx^{\mu} \hat{A}^{adj}_{\mu}  \right) {}^{a}_{c}
    \rho^{a}_{in} = \nonumber \\ = N^{-1} + 
    \rho^{b}_{in} 
    \mathcal{P} \exp \left( \int \limits_{\delta \gamma} dx^{\mu} \hat{A}^{adj}_{\mu} \right) {}^{a}_{b}
    \rho^{a}_{in}
\end{eqnarray}
where $\delta\gamma$ is the loop obtained by travelling forward in the loop $\gamma$ and backward in the loop $\gamma'$. 

For the purposes of further analysis we will apply the non-Abelian Stokes  theorem \cite{Fishbane:81, Diosi:83} to the expression (\ref{FidelityAsLoopIntegral}), as was proposed in the work \cite{Kuzmin:03}. In contrast to the work \cite{Kuzmin:03} we use adjoint representation of $su(N)$ algebra, thus we must first define several new objects. We define the curvature tensor $F^{a}_{\mu \nu}$ of the adiabatic connection as:
\begin{eqnarray}
\label{CurvatureDefinition}
F^{a}_{\mu \nu} (\lambda) = \frac{\partial A^{a}_{\mu} (\lambda)}{\partial \lambda^{\nu}} - \frac{\partial A^{a}_{\nu} (\lambda)}{\partial \lambda^{\mu}} - C^{a}_{\: bc} A_{\mu}^{b} (\lambda) A_{\nu}^{c} (\lambda)
\end{eqnarray}
The shifted curvature tensor $G^{a}_{\mu \nu}$ is defined as:
\begin{eqnarray}
\label{ShiftedCurvatureDefinition}
G^{a}_{\mu \nu} (\lambda, \lambda_{0}, \chi) = \mathcal{P} \exp \left( \int \limits_{\chi; \: \lambda}^{\lambda_{0}} d\lambda^{\mu} \hat{A}^{adj}_{\mu} \right) {}^{a}_{b} F^{b}_{\mu \nu} (\lambda)
\end{eqnarray}
where $\lambda_{0}$ is the so-called "reference point" and $\chi$ is some path which connects the points $\lambda_{0}$ and $\lambda$. The shifted curvature tensor is the curvature tensor parallel transported to the reference point. We also introduce the shifted curvature tensor in the adjoint representation $\left(\hat{G}^{adj}_{\mu \nu} \right){}^{a}_{c} = C^{a}_{\: bc} G^{b}_{\mu \nu}$.

Now it is possible to apply the non-Abelian Stokes theorem to the path-ordered exponent in (\ref{FidelityAsLoopIntegral}) and express the integral over the loop $\delta \gamma$ as the integral over the surface $S_{\delta \gamma}$ spanned on the loop $\delta \gamma$ \cite{Fishbane:81, Diosi:83}:
\begin{eqnarray}
\label{FidelityViaNAST}
f = N^{-1} + \rho^{b}_{in} 
    \mathcal{P}_{S_{\delta \gamma}} \exp \left( \int \limits_{S_{\delta \gamma}} d \sigma^{\mu \nu} \hat{G}^{adj}_{\mu \nu} \right) {}^{a}_{b}
    \rho^{a}_{in}
\end{eqnarray} 
where $d \sigma^{\mu \nu}$ is the element of area on $S_{\delta \gamma}$ and $\mathcal{P}_{S_{\delta \gamma}}$ is the path-ordering operator for the surface $S_{\delta \gamma}$. The initial point of the paths $\gamma'$ and $\gamma$ is chosen as the reference point $\lambda_{0}$ in the definition of the shifted curvature tensor (\ref{ShiftedCurvatureDefinition}). The surface $S_{\delta \gamma}$ can be chosen to be the surface of the minimal area spanned on the loop $\delta \gamma$. The paths $\chi$ are assumed to lie on the surface $S_{\delta \gamma}$. The choice of the paths $\chi$ completely determines the path-ordering operator for the surface $\mathcal{P}_{S_{\delta \gamma}}$ \cite{Fishbane:81, Diosi:83}.

The distance between the paths $\gamma'$ and $\gamma$ should be small, as otherwise HQC will lead to unpredictable results and will be therefore useless. Let us denote the difference between the coordinates of the corresponding points on the loops $\gamma'$ and $\gamma$ as $\delta \lambda^{\mu}$. Such definition implies that $\delta \lambda$ is the function of the points on the path $\gamma$. To the first order in $\delta \lambda^{\mu}$ we can write the element of area as $d \sigma^{\mu \nu} \approx d \lambda^{[ \mu} \delta \lambda^{\nu ]}$. We retain only the first-order terms in $\delta \lambda^{\mu}$ in the argument of the path-ordered exponent in (\ref{FidelityViaNAST}). However we do not expand the whole exponent to the terms of order $\delta \lambda^{\nu}$, as the total effect of numerous small errors can be quite significant for large loops $\gamma$, which can be inferred from the results of our numerical simulations (see further).  After these assumptions we can write, using the properties of the path-ordering operator $\mathcal{P}_{S_{\delta \gamma}}$ \cite{Fishbane:81, Diosi:83}:
\begin{eqnarray}
\label{IntegralApproximation}
  \mathcal{P}_{S_{\delta \gamma}} \exp \left( \int \limits_{S_{\delta \gamma}} d \sigma^{\mu \nu} \hat{G}^{adj}_{\mu \nu} \right) {}^{a}_{b} \approx \nonumber \\ 
\approx \mathcal{P} \exp \left( \int \limits_{\gamma} d \lambda^{\mu} \delta \lambda^{\nu} (\lambda) \hat{G}^{adj}_{\mu \nu} (\lambda) \right) {}^{a}_{b}
\end{eqnarray} 
Such approximation is valid if the shifted curvature tensor does not change significantly at distances of order $\delta \lambda$ and if $\delta \lambda$ is small compared to the loop size. Furthermore, to the first order in $d \lambda^{\mu}$ the path $\chi$ in (\ref{FidelityViaNAST}) can be assumed to coincide with the part of the path $\gamma$ which connects the initial point with the point where the shifted curvature tensor is calculated. 

Thus we obtain the following approximate expression for the fidelity:
\begin{eqnarray}
\label{FidelityFinal}
f \approx N^{-1} + \rho^{b}_{in} 
    \mathcal{P} \exp \left( \int \limits_{\gamma} d \lambda^{\mu} \delta \lambda^{\nu} \hat{G}^{adj}_{\mu \nu} \right) {}^{a}_{b}
    \rho^{a}_{in}
\end{eqnarray}

\section{\label{sec:RandomErrors} The fidelity in the case of random uncorrelated errors}

Errors in the values of external control parameters are unpredictable and essentially random, which makes the general expression (\ref{FidelityFinal}) for the fidelity quite impractical. Indeed, it is necessary to know the errors $\delta \lambda^{\mu}$ exactly in order to calculate the fidelity in the form (\ref{FidelityFinal}). One would like instead to predict the fidelity basing on the statistical properties of errors. This was done numerically in some particular cases \cite{Solinas:05}. Here we will obtain a general expression in the limit of small and uncorrelated errors.

As the fidelity is the probability of obtaining the correct result of quantum computations, the correct expression for the fidelity which takes into account the random character of errors is obtained by averaging the expression (\ref{FidelityFinal}) over all errors according to the multiplication law for conditional probability:
\begin{eqnarray}
\label{FidelityRandomDefinition}
 f \approx N^{-1} + \rho^{b}_{in} 
    \overline{\mathcal{P} \exp \left( \int \limits_{\gamma} d \lambda^{\mu} \delta \lambda^{\nu} \hat{G}^{adj}_{\mu \nu} \right) {}^{a}_{b}}
    \rho^{a}_{in}
\end{eqnarray}
where by the line over the expression we denote averaging with respect to errors in the values of external control parameters. In this work we will make the following assumptions concerning the statistical properties of errors:
\begin{enumerate}
    \item Expectation values of errors are equal to zero: $\overline{\delta \lambda^{\mu}} = 0$.
    \item Errors are statistically independent in the parts of the loop $\gamma$ separated by sufficiently large distance, so that one can introduce some characteristic correlation length $\lambda_{c}$ which has the meaning of the distance between uncorrelated errors and which determines how frequently the errors occur along the loop $\gamma$.
    \item Errors are small: $ || G^{a}_{\mu \nu} || \: \overline{|\delta \lambda^{\mu}|} \: \lambda_{c} \ll 1$, where the norm $|| G^{a}_{\mu \nu} || = \sqrt{ \sum \limits_{a, \: \mu, \: \nu} (G^{a}_{\mu \nu})^{2} }$. 
\end{enumerate}
It should be noted that $|| G^{a}_{\mu \nu} || = || F^{a}_{\mu \nu} ||$, and the smallness condition can be also written as:
\begin{eqnarray}
\label{GaussianDominanceCondition}
|| F^{a}_{\mu \nu} || \: \overline{|\delta \lambda^{\mu}|} \: \lambda_{c} \ll 1 
\end{eqnarray}

The expectation value of the path-ordered exponent in (\ref{FidelityRandomDefinition}) can be calculated using the van-Kampen cumulant expansion \cite{VanKampen:74, VanKampenStochasticProcesses}:
\begin{widetext}
\begin{eqnarray}
\label{PathExpVanKampenGeneral}
    \overline{\mathcal{P} \exp \left( \int \limits_{\gamma} d \lambda^{\mu} \delta \lambda^{\nu} \hat{G}^{adj}_{\mu \nu} \right)} = \nonumber \\ =
 \mathcal{P} \exp \left( \sum \limits_{k=1}^{\infty} 1/k! \int \limits_{\gamma} \ldots \int \limits_{\gamma} d \lambda^{\mu_{1}}_{1} \ldots d \lambda^{\mu_{k}}_{k}  \langle \langle \hat{G}^{adj}_{\mu_{1} \nu_{1}} (\lambda_{1}) \ldots \hat{G}^{adj}_{\mu_{k} \nu_{k}} (\lambda_{k}) \delta \lambda^{\nu_{1}} \ldots \delta \lambda^{\nu_{k}} \rangle \rangle \right)
\end{eqnarray}
\end{widetext}
where $\langle \langle \ldots \rangle \rangle$ denote the so-called cumulants \cite{VanKampen:74, VanKampenStochasticProcesses}. For instance the second-order cumulant is $\langle \langle A B \rangle \rangle = \overline{A B} - \overline{A} \: \overline{B}$. The summand with $k=1$ is equal to zero because the expectation value of errors is equal to zero. If the smallness condition (\ref{GaussianDominanceCondition}) holds, the contribution of high-order terms is small compared to the second-order terms, and they can be omitted in the van-Kampen expansion (\ref{PathExpVanKampenGeneral}), which leads to the following expression:
\begin{widetext}
\begin{eqnarray}
\label{PathExpVanKampen}
\overline{\mathcal{P} \exp \left( \int \limits_{\gamma} d \lambda^{\mu} \delta \lambda^{\nu} \hat{G}^{adj}_{\mu \nu} \right)} \approx \mathcal{P} \exp \left( 1/2 \: \int \limits_{\gamma} \int \limits_{\gamma} 
d \lambda_{1}^{\mu} d \lambda_{2}^{\nu} \overline{\delta \lambda^{\alpha} (\lambda_{1}) \delta \lambda^{\beta} (\lambda_{2})}
\hat{G}^{adj}_{\mu \alpha} (\lambda_{1}) \hat{G}^{adj}_{\nu \beta} (\lambda_{2}) \right)
\end{eqnarray}
\end{widetext}
As the correlation function $\overline{\delta \lambda^{\alpha} (\lambda_{1}) \delta \lambda^{\beta} (\lambda_{2})}$ is not small only for close values of $\lambda_{1}$ and $\lambda_{2}$, that is only for close enough points on the loop $\gamma$, it is possible to integrate over $\lambda_{1}$ in (\ref{PathExpVanKampen}). The integrand is large enough only in the vicinity of the size $\lambda_{c}$ of the point $\lambda_{2}$, thus the path-ordered exponent in (\ref{PathExpVanKampen}) can be approximately calculated as:
\begin{eqnarray}
\label{PathExpVanKampenSimple}
    \overline{\mathcal{P} \exp \left( \int \limits_{\gamma} d \lambda^{\mu} \delta \lambda^{\nu} \hat{G}^{adj}_{\mu \nu} \right)} \approx \nonumber \\ \approx \mathcal{P} \exp \left( 1/2 \: \lambda_{c} \int \limits_{\gamma} d \lambda^{\mu}
    \tau^{\nu} \overline{\delta \lambda^{\alpha} \delta \lambda^{\beta}} \hat{G}^{adj}_{\mu \alpha} \hat{G}^{adj}_{\nu \beta} \right)
\end{eqnarray}
where $\tau^{\nu}$ is the tangent vector of the loop $\gamma$. Note that such integration can be regarded as the implicit definition of the correlation length $\lambda_{c}$:
\begin{eqnarray}
\label{CorrLengthDefinition}
 \int \limits_{\gamma} \int \limits_{\gamma} d \lambda_{1}^{\mu} d \lambda_{2}^{\nu} \overline{\delta \lambda^{\alpha} (\lambda_{1}) \delta \lambda^{\beta} (\lambda_{2})} \approx \nonumber \\ \approx
\lambda_{c} \int \limits_{\gamma} d \lambda^{\mu} \tau^{\nu}(\lambda) \overline{\delta \lambda^{\alpha} (\lambda) \delta \lambda^{\beta} (\lambda)}
\end{eqnarray}

After applying the Van-Kampen expansion the final approximate expression for the fidelity reads:
\begin{eqnarray}
\label{FidelityFinalRandom}
 f \approx N^{-1} + \nonumber \\ + \rho^{b}_{in} 
    \mathcal{P} \exp \left( 1/2 \: \lambda_{c} \int \limits_{\gamma} d \lambda^{\mu}
    \tau^{\nu} \overline{\delta \lambda^{\alpha} \delta \lambda^{\beta}} \hat{G}^{adj}_{\mu \alpha} \hat{G}^{adj}_{\nu \beta} \right) {}^{a}_{b}
    \rho^{a}_{in}
\end{eqnarray}

The expression (\ref{FidelityFinalRandom}) shows that the stability of quantum computation essentially depends on:
\begin{enumerate}
  \item The duration and closeness of errors (the factor $\lambda_{c}$ in (\ref{FidelityFinalRandom}))
  \item The shifted curvature tensor on the space of external parameters $\hat{G}^{adj}_{\mu\nu}$
  \item The expectation value of the square of the magnitude of errors
  \item The length of the contour $\gamma$ in the space of external control parameters
\end{enumerate}
The fidelity should in general decay with the length of the loop $\gamma$. Fidelity decay rate is proportional to the expectation value of the square of errors and the square of the shifted curvature tensor.

\section{\label{sec:Simulations}Results of numerical simulations}

Numerical simulations were performed in order to verify the theoretical results of the previous section. We used two-dimensional Hilbert space of the quantum register and worked with generators of $SU(2)$ group. Normalized generators of $SU(2)$ group are $\hat{T}_{a} = \hat{\sigma}_{a}/\sqrt{2}$, $a = 1,2,3$, where $\hat{\sigma}_{a}$ are the Pauli matrices, and the structure constants of $su(2)$ Lie algebra are $C^{c}_{ab} = \epsilon_{abc}/\sqrt{2}$. Parametric space was also assumed to be two-dimensional with parameters $\lambda^{1} = x$, $\lambda^{2} = y$. We have chosen a very simple form for the adiabatic connection, which however yields nontrivial coordinate dependence for the shifted curvature tensor: 
\begin{eqnarray}
\label{SimulationConventions1}
 \hat{A}_{x} = \hat{\sigma}_{1}, \quad
 \hat{A}_{y} = \hat{\sigma}_{2}, \quad
 A^{1}_{x} = \sqrt{2}, \quad A^{2}_{y} = \sqrt{2} \nonumber \\
 F^{3}_{xy} = - F^{3}_{yx} = 2 \sqrt{2}
\end{eqnarray}
with all other components being equal to zero. As both $\hat{A}^{adj}_{x}$ and $\hat{A}^{adj}_{y}$ do not commute with $\hat{F}^{adj}_{xy}$, the shifted curvature tensor is not constant. Such adiabatic connection is, for example, approximately equal to the adiabatic connection for the optical HQC based on Kerr medium (described in the section IV of the work \cite{Pachos:01}) at small values of displacement parameters ($x$ and $y$) and zero squeezing parameter (denoted as $r_{1}$ in the work \cite{Pachos:01}). We used square loop $\gamma$ of the size $L \times L$. The components of the initial density matrix of the quantum register were $\rho_{11}= 1, \rho_{12} = \rho_{21} = \rho_{22}= 0$.

Errors were simulated in the following way: a set of $N_{err}$ points on the loop $\gamma$ was chosen, then each of these points was shifted independently from the others by a gaussian-distributed random vector. To obtain the loop $\gamma'$ these shifted points were connected by a cubic spline. Distribution of errors was assumed to be isotropic, i.e. $\overline{\delta \lambda^{\alpha} \delta \lambda^{\beta}} \sim \delta_{\alpha \beta}$. Parallel transport equation (\ref{StateParallelTransport}) was solved using the fourth-order Runge-Kuttha method. The value of the fidelity was calculated using the definition (\ref{FidelityDefinition}). Finally the fidelity was averaged over the implementations of errors. Theoretical value of the fidelity (\ref{FidelityFinalRandom}) was also calculated numerically. It should be noted that we have used not the correlation length $\lambda_{c}$ but the number of error points $N_{err}$ on the loop as the parameter in our simulations. Correlation length can be estimated as the length of the loop divided by the number of error points $N_{err}$: $\lambda_{c} \approx 4 L N^{-1}_{err}$. When calculating the fidelity as the function of the loop length, $N_{err}$ was increased proportionally to $L$, so that the correlation length remained constant.

\begin{figure}[h]
  \includegraphics[width=8cm]{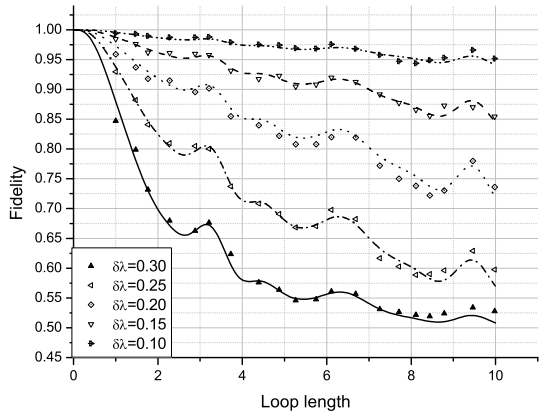}\\
  \caption{The fidelity as the function of the loop size $L$ for different magnitudes of errors at fixed correlation length ($\lambda_{c} = 0.1$).}
  \label{FidelityVsLength}
\end{figure}

\begin{figure}[h]
  \includegraphics[width=8cm]{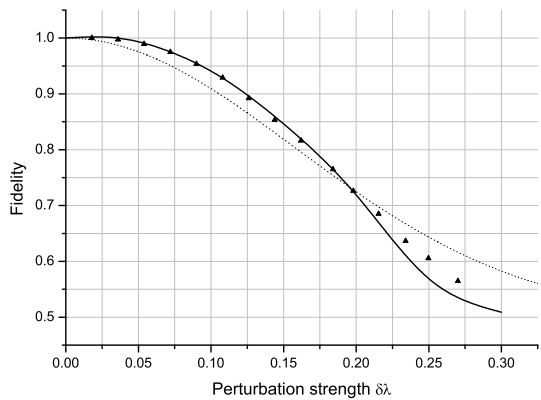}\\
  \caption{The fidelity as the function of the magnitude of errors $\sqrt{\overline{\delta \lambda^{2}}}$ at fixed loop size ($L = 10$) and correlation length ($\lambda_{c} = 0.1$).}
  \label{FidelityVsPerturbation}
\end{figure}

\begin{figure}[h]
  \includegraphics[width=8cm]{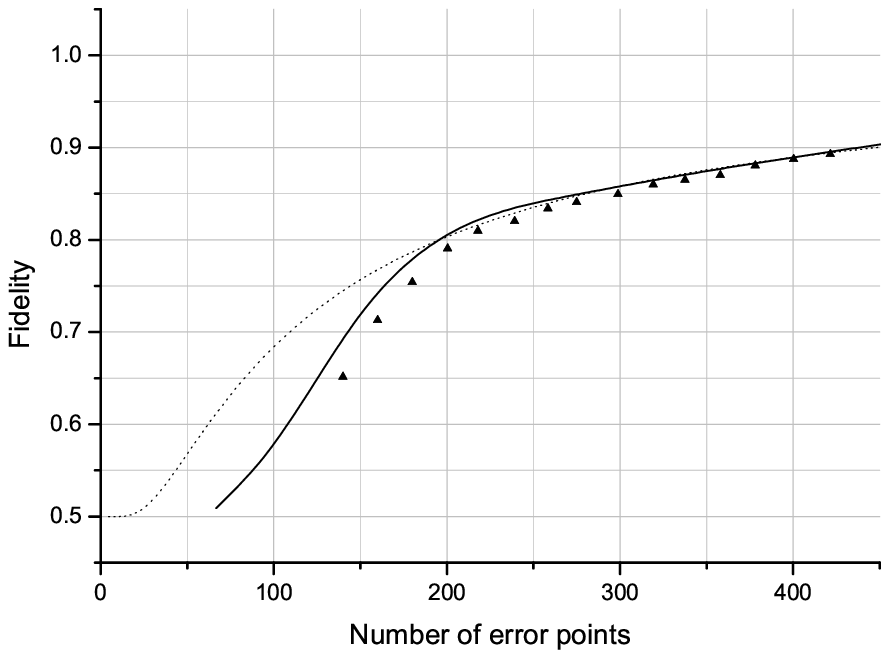}\\
  \caption{The fidelity as the function of the number of error points $N_{err}$ at fixed loop size ($L = 10$) and perturbation strength ($\delta \lambda = 0.15$).}
  \label{FidelityVsLc}
\end{figure}

As can be inferred from the figures \ref{FidelityVsLength}, \ref{FidelityVsPerturbation} and \ref{FidelityVsLc}, the results of the simulations are in a good agreement with the expression (\ref{FidelityFinalRandom}). Theoretical values of the fidelity obtained using the expression (\ref{FidelityFinalRandom}) are plotted with solid lines. It is worth noting that the expression (\ref{FidelityFinalRandom}) involves matrix exponents rather than just scalars, and thus the fidelity decay in general differs from simple exponential. In order to emphasize this difference exponential fits of the simulation results were also plotted on the figures \ref{FidelityVsPerturbation} and \ref{FidelityVsLc}. In figure \ref{FidelityVsPerturbation} the fitting function is $0.5 + 0.5 \exp{(- \alpha \delta \lambda^{2} )}$, which is a naive estimation for (\ref{FidelityFinalRandom}) if one treats the exponent as a scalar. Thus the dependence of the fidelity on perturbation strength differs from gaussian. Similarly, dotted line on the figure \ref{FidelityVsPerturbation} corresponds to the function $0.5 + 0.5 \exp{(- \beta / N_{err}  )}$. Here the exponential approximation is also not very good. 

Figures (\ref{FidelityVsLc}) and (\ref{FidelityVsPerturbation}) show that there are small discrepancies between theoretical results and the results of the simulations at large correlation lengths $\lambda_{c}$ and at large error magnitudes, which means that the value of (\ref{GaussianDominanceCondition}) is not sufficiently small in this region, and the van-Kampen expansion with second-order terms can not be applied to calculate the expectation value of path-ordered exponent in (\ref{PathExpVanKampen}). Let us estimate the smallness parameter for the extremal values of perturbation strength and correlation length used in our simulations. The norm of the curvature tensor is $|| F^{a}_{\mu \nu} || = \sqrt{ \sum \limits_{a, \: \mu, \: \nu} (F^{a}_{\mu \nu})^{2} } = 4 $. When $\delta \lambda = 0.3$, $L=10$, $N_{err} = 400$, $\lambda_{c} \approx 0.1$, the smallness parameter is $|| F^{a}_{\mu \nu} || \: \delta \lambda \: \lambda_{c} \approx 0.12$ (figures \ref{FidelityVsLength} and \ref{FidelityVsPerturbation}). When $\delta \lambda = 0.15$, $L=10$, $N_{err} = 100$, $\lambda_{c} \approx 0.4$, the smallness parameter is $|| F^{a}_{\mu \nu} || \: \delta \lambda \: \lambda_{c} \approx 0.24$ (figure \ref{FidelityVsLc}). Thus the results of the simulations indicate that the expression (\ref{FidelityFinalRandom}) is valid even if the fidelity is not close to unity, that is when the total effect of numerous small errors is not small. The results of our simulations are in agreement with those in \cite{Solinas:05}.

\section{\label{sec:Conclusions} Conclusions}

In this paper we investigated the influence of random uncorrelated errors on the stability of holonomic quantum computations in the most general case, without explicitly specifying the adiabatic connection or the loop in the space of external control parameters. A simple expression for the fidelity was obtained, which is valid when the condition $|| F^{a}_{\mu \nu} || \: \overline{|\delta \lambda^{\mu}|} \: \lambda_{c} \ll 1$ holds, where $|| F^{a}_{\mu \nu} ||$ is the norm of the curvature tensor of the adiabatic connection. This expression can be used to predict fidelity decay rates by specifying just a few parameters such as the average square of parameter deviations and the frequency of errors. It should be noted that though the final expression is valid for small magnitude of errors $\overline{|\delta \lambda^{\mu}|}$, it is applicable even when the fidelity is not close to unity. Our theoretical results describe with a good precision the data obtained by numerical simulations, some discrepancies arising only when the condition (\ref{GaussianDominanceCondition}) does not hold. 


\begin{thebibliography}{17}
\expandafter\ifx\csname natexlab\endcsname\relax\def\natexlab#1{#1}\fi
\expandafter\ifx\csname bibnamefont\endcsname\relax
  \def\bibnamefont#1{#1}\fi
\expandafter\ifx\csname bibfnamefont\endcsname\relax
  \def\bibfnamefont#1{#1}\fi
\expandafter\ifx\csname citenamefont\endcsname\relax
  \def\citenamefont#1{#1}\fi
\expandafter\ifx\csname url\endcsname\relax
  \def\url#1{\texttt{#1}}\fi
\expandafter\ifx\csname urlprefix\endcsname\relax\def\urlprefix{URL }\fi
\providecommand{\bibinfo}[2]{#2}
\providecommand{\eprint}[2][]{\url{#2}}

\bibitem[{\citenamefont{Zanardi and Rasetti}(1999)}]{Zanardi:99}
\bibinfo{author}{\bibfnamefont{P.}~\bibnamefont{Zanardi}} \bibnamefont{and}
  \bibinfo{author}{\bibfnamefont{M.}~\bibnamefont{Rasetti}},
  \bibinfo{journal}{Physics Letters A} \textbf{\bibinfo{volume}{264}},
  \bibinfo{pages}{94} (\bibinfo{year}{1999}).

\bibitem[{\citenamefont{Pachos}(2002)}]{Pachos:02}
\bibinfo{author}{\bibfnamefont{J.}~\bibnamefont{Pachos}},
  \bibinfo{journal}{Physical Review A} \textbf{\bibinfo{volume}{66}},
  \bibinfo{pages}{042318} (\bibinfo{year}{2002}).

\bibitem[{\citenamefont{Pachos et~al.}(2000)\citenamefont{Pachos, Zanardi, and
  Rasetti}}]{Pachos:05}
\bibinfo{author}{\bibfnamefont{J.}~\bibnamefont{Pachos}},
  \bibinfo{author}{\bibfnamefont{P.}~\bibnamefont{Zanardi}}, \bibnamefont{and}
  \bibinfo{author}{\bibfnamefont{M.}~\bibnamefont{Rasetti}},
  \bibinfo{journal}{Physical Review A} \textbf{\bibinfo{volume}{61}},
  \bibinfo{pages}{010305(R)} (\bibinfo{year}{1999}).

\bibitem[{\citenamefont{Pachos and Chountasis}(2000)}]{Pachos:00}
\bibinfo{author}{\bibfnamefont{J.}~\bibnamefont{Pachos}} \bibnamefont{and}
  \bibinfo{author}{\bibfnamefont{S.}~\bibnamefont{Chountasis}},
  \bibinfo{journal}{Physical Review A} \textbf{\bibinfo{volume}{62}},
  \bibinfo{pages}{052318} (\bibinfo{year}{2000}).

\bibitem[{\citenamefont{Kuvshinov and Kuzmin}(2003)}]{Kuzmin:03}
\bibinfo{author}{\bibfnamefont{V.~I.} \bibnamefont{Kuvshinov}}
  \bibnamefont{and} \bibinfo{author}{\bibfnamefont{A.~V.}
  \bibnamefont{Kuzmin}}, \bibinfo{journal}{Physics Letters A}
  \textbf{\bibinfo{volume}{316}}, \bibinfo{pages}{391 } (\bibinfo{year}{2003}).

\bibitem[{\citenamefont{Solinas et~al.}(2004)\citenamefont{Solinas, Zanardi,
  and Zanghi}}]{Solinas:05}
\bibinfo{author}{\bibfnamefont{P.}~\bibnamefont{Solinas}},
  \bibinfo{author}{\bibfnamefont{P.}~\bibnamefont{Zanardi}}, \bibnamefont{and}
  \bibinfo{author}{\bibfnamefont{N.}~\bibnamefont{Zanghi}},
  \bibinfo{journal}{Physical Review A} \textbf{\bibinfo{volume}{70}},
  \bibinfo{pages}{042316} (\bibinfo{year}{2004}).

\bibitem[{\citenamefont{Carollo et~al.}(2003)\citenamefont{Carollo,
  Fuentes-Guridi, Santos, and Vedral}}]{Carollo:03}
\bibinfo{author}{\bibfnamefont{A.}~\bibnamefont{Carollo}},
  \bibinfo{author}{\bibfnamefont{I.}~\bibnamefont{Fuentes-Guridi}},
  \bibinfo{author}{\bibfnamefont{M.~F.} \bibnamefont{Santos}},
  \bibnamefont{and} \bibinfo{author}{\bibfnamefont{V.}~\bibnamefont{Vedral}},
  \bibinfo{journal}{Physical Review Letters} \textbf{\bibinfo{volume}{90}},
  \bibinfo{pages}{160402} (\bibinfo{year}{2003}).

\bibitem[{\citenamefont{Chiara and Palma}(2003)}]{DeChiara:03}
\bibinfo{author}{\bibfnamefont{G.} \bibnamefont{DeChiara}} \bibnamefont{and}
  \bibinfo{author}{\bibfnamefont{G.~M.} \bibnamefont{Palma}},
  \bibinfo{journal}{Physical Review Letters} \textbf{\bibinfo{volume}{91}},
  \bibinfo{pages}{090404} (\bibinfo{year}{2003}).

\bibitem[{\citenamefont{Kuvshinov and Kuzmin}(2005)}]{Kuzmin:05}
\bibinfo{author}{\bibfnamefont{V.~I.} \bibnamefont{Kuvshinov}}
  \bibnamefont{and} \bibinfo{author}{\bibfnamefont{A.~V.}
  \bibnamefont{Kuzmin}}, \bibinfo{journal}{Physics Letters A}
  \textbf{\bibinfo{volume}{341}}, \bibinfo{pages}{450 } (\bibinfo{year}{2005}).

\bibitem[{\citenamefont{Wilczek and Zee}(1984)}]{Wilczek:84}
\bibinfo{author}{\bibfnamefont{F.}~\bibnamefont{Wilczek}} \bibnamefont{and}
  \bibinfo{author}{\bibfnamefont{A.}~\bibnamefont{Zee}},
  \bibinfo{journal}{Physical Review Letters} \textbf{\bibinfo{volume}{52}},
  \bibinfo{pages}{2111 } (\bibinfo{year}{1984}).

\bibitem[{\citenamefont{Peres}(1984)}]{Peres:84}
\bibinfo{author}{\bibfnamefont{A.}~\bibnamefont{Peres}},
  \bibinfo{journal}{Physical Review A} \textbf{\bibinfo{volume}{30}},
  \bibinfo{pages}{1610 } (\bibinfo{year}{1984}).

\bibitem[{\citenamefont{Prosen and \v{Z}nidari\v{c}}(2001)}]{Prozen:01}
\bibinfo{author}{\bibfnamefont{T.}~\bibnamefont{Prosen}} \bibnamefont{and}
  \bibinfo{author}{\bibfnamefont{M.}~\bibnamefont{\v{Z}nidari\v{c}}},
  \bibinfo{journal}{Journal of Physics A} \textbf{\bibinfo{volume}{34}},
  \bibinfo{pages}{L681} (\bibinfo{year}{2001}).

\bibitem[{\citenamefont{Fishbane et~al.}(1981)\citenamefont{Fishbane,
  Gasiorowicz, and Kaus}}]{Fishbane:81}
\bibinfo{author}{\bibfnamefont{P.}~\bibnamefont{Fishbane}},
  \bibinfo{author}{\bibfnamefont{S.}~\bibnamefont{Gasiorowicz}},
  \bibnamefont{and} \bibinfo{author}{\bibfnamefont{P.}~\bibnamefont{Kaus}},
  \bibinfo{journal}{Physical Review D} \textbf{\bibinfo{volume}{24}},
  \bibinfo{pages}{2324} (\bibinfo{year}{1981}).

\bibitem[{\citenamefont{Diosi}(1983)}]{Diosi:83}
\bibinfo{author}{\bibfnamefont{L.}~\bibnamefont{Diosi}},
  \bibinfo{journal}{Physical Review D} \textbf{\bibinfo{volume}{27}},
  \bibinfo{pages}{2552} (\bibinfo{year}{1983}).

\bibitem[{\citenamefont{van Kampen}(1974)}]{VanKampen:74}
\bibinfo{author}{\bibfnamefont{N.~G.} \bibnamefont{van Kampen}},
  \bibinfo{journal}{Physica} \textbf{\bibinfo{volume}{74}}, \bibinfo{pages}{215
  } (\bibinfo{year}{1974}).

\bibitem[{\citenamefont{van Kampen}(1981)}]{VanKampenStochasticProcesses}
\bibinfo{author}{\bibfnamefont{N.~G.} \bibnamefont{van Kampen}},
  \emph{\bibinfo{title}{Stochastic processes in physics and chemistry}}
  (\bibinfo{publisher}{North-Holland}, \bibinfo{address}{Amsterdam},
  \bibinfo{year}{1981}).

\bibitem[{\citenamefont{Pachos and Zanardi}(2001)}]{Pachos:01}
\bibinfo{author}{\bibfnamefont{J.}~\bibnamefont{Pachos}} \bibnamefont{and}
  \bibinfo{author}{\bibfnamefont{P.}~\bibnamefont{Zanardi}},
  \bibinfo{journal}{International Journal of Modern Physics B}
  \textbf{\bibinfo{volume}{15}}, \bibinfo{pages}{1257} (\bibinfo{year}{2001}).

\end{thebibliography}

\end{document}